# STUDY ON PLANAR WHISPERING GALLERY DIELECTRIC RESONATORS. II. A MULTIPLE-BAND DEVICE

Giuseppe Annino[@], Mario Cassettari, and Massimo Martinelli

Istituto di Fisica Atomica e Molecolare[*], CNR, Area della Ricerca, via Moruzzi 1, 56124 Pisa, Italy.

[*] now Istituto per i Processi Chimico-Fisici

[@] Corresponding author.           E-mail: geannino@ifam.pi.cnr.it

(March 05, 2002)

## ABSTRACT

The basic theory underlying the realization of simple multiple-band non-homogeneous dielectric resonators, whose spectral response is the overlap of single-resonator frequency bands, is developed exploiting a general approach discussed in the previous companion paper. The limit frequencies of the proposed devices, given only by the dielectric properties of the involved materials, can differ in principle by several decades. Experimental confirmations have been obtained on a composite structure built up with teflon and polyethylene; as predicted by the theory, the overall band includes frequencies which range about from 20 GHz to more than 400 GHz, when high frequency resonances are selectively excited. The localization of the higher frequency radiation between the positive steps of the dielectric constant, which is the basic properties of these non-homogeneous resonators, has been experimentally verified by mapping the electromagnetic field intensity. Possible applications of multiple-band Whispering Gallery dielectric resonators are finally outlined.

## 1. Introduction

Whispering gallery dielectric resonators (WGDRs) form a class of rotationally invariant resonant structures, typically characterized by high values of the azimuthal index **n**; they exhibit then a multiplicity of sharp resonances and can be considered wideband resonators since their frequency band is in general much greater than the free spectral range. WGDR appears as a natural candidate for the realization of an ultra-wideband resonant structure, in addition to the broad field of applications that these simple resonators have been applied to **[1-7]**.

The limit properties of planar homogeneous WGDRs have been discussed in the companion paper **[8]**, with particular regard to their frequency response; although very wide in principle, the obtainable frequency band is severely limited by the presence of transverse modes, that in a homogeneous structure behave as the fundamental ones. The selective excitation of a specific family of mode is then quite difficult, so that the effective frequency band (ranging from $\mathbf{w}_{low}$ to $\mathbf{w}_{upp}$, see Ref. **[8]**) is limited at least by a factor 3 in terms of decades in comparison to the ideal frequency band (ranging from $\mathbf{w}_{min}$ to $\mathbf{w}_{max}$), calculated for fundamental modes only. Moreover, the realization of a homogeneous ultra-wideband WGDR is limited by practical reasons, owing to the difficulty to find a dielectric material transparent over the whole ideal band. A solution that overcomes the intrinsic limitations of a homogeneous WGDR is obtained coupling different WGDRs, taking advantage from their open structure, in which no metallic shields are necessary. The inhomogeneity in the dielectric properties of the composite structure induces a differentiation in the properties of their resonance modes, that can be in turn selectively excited. As a consequence, the spectral response of a composite WGDR results the overlap of the spectral responses of the employed components. In order to limit the irradiation losses, a geometry preserving the cylindrical symmetry is preferable; by this way the scattering of radiation due to the inhomogeneity along the azimuthal direction is avoided. A simple composite structure that meets these requirements is represented by the already studied stacked configuration **[9, 10]**. The analysis of the working principles of basic stacked configurations together with a preliminary experimental characterization represent the main aim of this paper.

In Sect. 2 generalities about WGDRs are recalled, and the structures under investigation are introduced. Sect. 3 deals with the working principles of a partially coupled composite WGDR; in particular it is shown how its minimum and maximum working frequencies can reach the ideal limits proper of the involved dielectric materials. The partially coupled configuration represents the starting point for the analysis of the fully coupled WGDR, studied in Sect. 4. In subsection 4a it is demonstrated that this configuration can reach the high frequency ideal limit only by a selective excitation of resonances; possible selection mechanisms among modes are then discussed in subsection 4b. In Sect. 5 a fully coupled WGDR built up with teflon and polyethylene is designed and experimentally characterized; the results obtained in the low frequency region are discussed in subsection 5a, while the results concerning the high frequency region are discussed in subsection 5b. Finally, possible applications are suggested in Sect. 6, together with some concluding remarks.

## 2. Generalities

The basic information about the resonance modes of a WGDR concerns the polarization and the spatial behavior of their electromagnetic fields. The modes are typically collected in two different families, whose elements are labeled as $WGE_{n,m,l}$ and $WGH_{n,m,l}$; the WGE family includes quasi-transverse electric modes, while the WGH family includes quasi-transverse magnetic modes. The modal indices have the following meaning: **n**, the azimuthal index, gives the number of wavelengths in a whole turn around the resonator; **m**, the radial index, gives the number of nodes of the energy flux along the radius of the resonator, and **l**, the axial index, gives the number of nodes of the energy flux along the axis of the resonator. The same nomenclature holds for resonators obtained from the stacked composition of different planar WGDRs, whose simpler implementations are reported in Fig. 1. In particular, Fig. 1a depicts a "partially coupled" stacked WGDR, while Fig. 1b depicts a "fully coupled" stacked WGDR. In the following, the real dielectric constant and the loss angle of the central disc will be indicated by $\mathbf{e}_1$ and $\mathbf{d}_1$, respectively, while the analogous parameters of the two identical outer discs will be indicated by $\mathbf{e}_2$ and $\mathbf{d}_2$, respectively. The whole structure is assumed to be embedded in an external medium having $\mathbf{e}_{ext} < \mathbf{e}_1, \mathbf{e}_2$ and negligible losses.

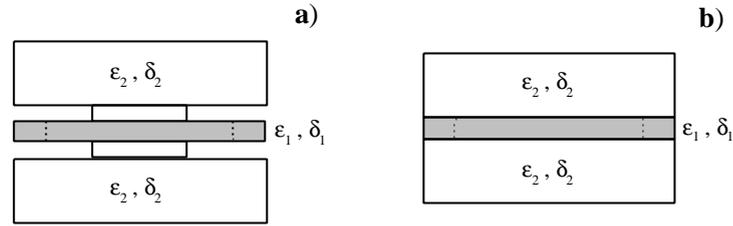

*Fig. 1.* *Composite planar whispering gallery dielectric resonators.*
 *a) Partially coupled configuration.   b) Fully coupled configuration.*
 *Shaded areas represent the high frequency regions.*

The analysis of the composite WGDR will be focused on the behavior of transverse axial modes ($l \neq 0$, $m = 0$); high frequency resonances with $m \neq 0$ can be suppressed making a central hole in the inner disc [11], while low frequency radial modes can be suppressed with a suitable choice of the geometry [8].

### 3. Partially coupled multiple-band WGDR

Let us first consider the arrangement of Fig. 1a, where the outer discs are not in contact with the central one. With a suitable choice of the thickness of the different regions, the central disc can be decoupled from the lateral ones in a proper high frequency band, while it can be strongly coupled in a proper low frequency band. With these assumptions, the central region behaves as an isolated resonator and the ideal limit $\omega_{max}$, fixed by its dielectric losses, can be reached. If the high frequency band is assumed centered around $\omega_{HF}$ ($\leq \omega_{max}$), the thickness of this region will results of the order of the wavelength in the medium $l_{HF} = \lambda(\omega_{HF})$ [8]. Analogously, the inner resonator can be decoupled from the outer discs if the gap between the central region and the lateral ones is of the order of $l_{HF}$ or greater.

On the other side, the whole structure behaves as a single resonator in the frequency band around $\omega_{LF}$, provided that: **a)** its total thickness is of the order of $l_{LF} = \lambda(\omega_{LF})$, and **b)** the dielectric inhomogeneity introduces weak perturbations in the modes of a resonator having identical overall dimensions and homogeneous dielectric constant $\varepsilon_2$. This latter requirement, equivalent to $l_{LF} \gg l_{HF}$, holds when $l_{LF} > 10\, l_{HF}$, as verified by the theoretical approach discussed in [10]. The spectral response of this composite resonator can be then given by the overlap between the frequency band of the inner resonator and the frequency band of the whole (homogeneous) structure, the minimum separation between these bands being given by the previous inequality.

The materials which form the different parts of a partially coupled resonator can be chosen without any constraint on their dielectric properties (provided that the consistence condition $\omega_{max} > \omega_{min}$ is satisfied in each region). When the employed dielectric constants are different, the considerations made in [8] hold in a generalized form. The limit frequencies for fundamental modes are obtained, analogously to the procedure of Sect. 3 in Ref [8], from the equations

$$\left(\frac{r}{l_2}\right)_{min} = \frac{\sqrt{\varepsilon_2}}{\sqrt{\varepsilon_2} - \sqrt{\varepsilon_{ext}}} \quad (1)$$

and

$$\left(\frac{r}{l_1}\right)_{max} = \frac{1}{2\pi \tan \delta_1} + 1 \quad (2),$$

where $r$ is the radius of the composite resonator. Eqs. 1 and 2 lead to

$$\frac{\omega_{max}}{\omega_{min}} = \left(\frac{1}{2\pi \tan \delta_1} + 1\right) \frac{\sqrt{\varepsilon_2} - \sqrt{\varepsilon_{ext}}}{\sqrt{\varepsilon_1}} \quad (3);$$

the limit frequencies of the proposed device can then differ by several decades.

The **Q** merit factor of a composite WGDR can be expressed in terms of the involved loss angles by means of the following general expression, which holds when irradiation losses are negligible

$$Q^{-1} = \eta_1 \tan \delta_1 + \eta_2 \tan \delta_2 \quad (4);$$

here $\eta_1$ is the fraction of the e.m. energy stored in the inner disc and $\eta_2$ is the fraction stored in the outer discs. The **Q** of the high frequency resonances ($\eta_1 \sim 1$) is limited by the dielectric losses of the inner region. For low frequency resonances having $l=0$, $\eta_1$ is of the order of the ratio between the thickness of the central disc and the thickness of the whole resonator, that is $\eta_1 \sim \frac{l_{HF}}{l_{LF}}$; the relative **Q** factor is then limited by the dielectric losses of the outer discs, provided that the condition $\tan \delta_1 < \frac{l_{LF}}{l_{HF}} \tan \delta_2$ holds.

## 4. Fully coupled multiple-band WGDR

The structure shown in Fig. 1a can be modified into a fully coupled WGDR, represented in Fig 1b; the working principles of the latter configuration are more complicated in comparison to those of a partially coupled WGDR, since the different parts are now coupled even in the high frequency region.

### *4a. Working principles*

A qualitative but physically transparent explanation of the working principles of fully coupled resonators can be obtained in terms of the geometrical optics representation, that applies to WGDRs under general conditions **[12]**. When $\varepsilon_1 > \varepsilon_2$ the modes that are not confined in the inner region propagate in the outer discs with lower axial propagation constant, analogously to the plane waves refraction **[10]**. In general, some of these modes will be confined by the outer plane surfaces of the structure, while the remaining ones will be axially irradiated. As a consequence, modes with lower axial indices (i.e. with lower axial propagation constants) can be confined in the inner disc, modes with intermediate axial indices are distributed in the whole structure, while modes with higher axial indices are axially irradiated. If on the contrary $\varepsilon_1 \leq \varepsilon_2$, the total reflection cannot occur in the inner region and all modes are distributed over the whole structure.

The previous considerations can be formalized in the framework of the theoretical approach discussed in **[10]**, by which resonance frequencies and fields distribution for each resonance mode can be calculated. As an illustrative example, calculations have been performed around 240 GHz on a structure formed by a central disc 1 mm thick and 16 mm in diameter, having $\varepsilon_1$=4.5, placed between two equal discs 5 mm thick and having $\varepsilon_2$=2. The resonances lying inside a free spectral range of the fundamental modes are compared with those of a homogeneous resonator with identical overall dimensions and $\varepsilon$=2. For sake of simplicity only modes of the form $WGH_{n,0,1}$ were investigated; similar results hold for the WGE family. The behavior of the modes can be conveniently discussed in terms of their tangential phase velocity $v_j(r)$, here calculated in correspondence of the rim of the resonator, where the excitation is typically localized; it results that **[12]**

$$v_j(r) = \frac{\omega \cdot r}{n} = \frac{\omega \cdot r}{\beta \times r_m} \stackrel{def}{=} \hat{v}_j \qquad (5),$$

where $r_m$ is the so-called modal caustic and $\beta$ is the transverse propagation constant (calculated on $r_m$) **[8]**.

Fig. 2 reports $\hat{v}_\varphi$ for the homogeneous and the inhomogeneous composite WGDR as a function of the axial index **l**. The number of allowed modes is greater in the composite structure, consistently with its greater optical thickness. For the homogeneous resonator, $\hat{v}_\varphi$ is practically constant for low axial indices, and increases in a regular manner with **l**, approaching and also exceeding the speed of light in the vacuum **c**; this effect is due to the increase of the axial propagation constant with **l** and to the related decrease of **b** [12]. On the contrary, for the composite structure $\hat{v}_\varphi$ shows near l=3 a sharp transition, which corresponds to the crossing from modes confined in the central region to modes distributed over the whole structure, as also verified from their fields distribution.

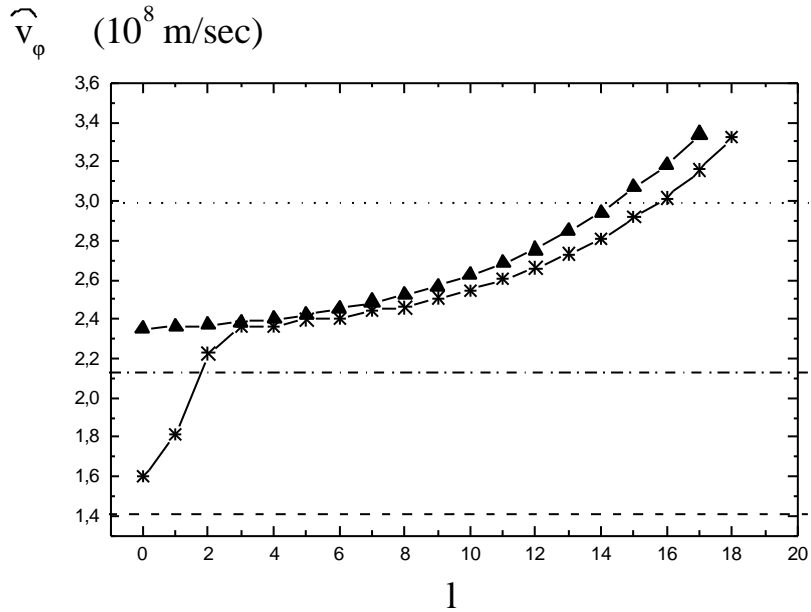

$\hat{v}_\varphi$ $(10^8$ m/sec$)$

*Fig. 2 Phase velocity near the rim of the resonator vs. the allowed axial indices. Triangles refer to the homogeneous resonator, stars to the composite resonator. Dotted line indicates the phase velocity of a plane wave in vacuum, dotted-dashed line indicates the phase velocity in a medium having **e**=2, and dashed line indicates the phase velocity in a medium having **e**=4.5.*

For the mode with **l=0,** $\hat{v}_\varphi$ is close to $\frac{c}{\sqrt{e_1}}$ ; modes with **l**=1 and **l**=2 are still confined in the inner disc but, due to the increasing axial propagation constant, the fraction of their energy distributed outside the central region becomes relevant; the resulting phase velocity is thus pulled towards the phase velocity proper of the outer discs. Finally, starting from $l=3$ the modes are distributed on the whole structure, so their phase velocity is close to $\frac{c}{\sqrt{\varepsilon_2}}$, analogously to the case of homogeneous resonator. The transition between modes confined in the inner disc and modes distributed on the whole structure can be regulated varying the thickness of the inner region; in particular, for a suitably thin inner disc only the mode with $l=0$ is confined. Similar results hold for modes with index $m \neq 0$.

As a conclusion, in the fully coupled WGDR the presence of transverse modes is intrinsic of the structure, so that the achievement of the maximum working frequency necessarily requires a selective excitation of modes confined in the inner region. Under this condition, the behavior of a fully coupled WGDR simply reduces to that of a partially coupled WGDR.

The following subsection is devoted to the analysis of the e.m. excitation of non-homogeneous WGDRs.

### *4b. Selective excitation of modes*

The coupling of e.m. radiation from a guiding structure to a WGDR can be analyzed in general way by means of the coupled mode theory **[13, 14]**. In this approach the fraction of power transferred to the resonator is given by the product of two factors: an overlap integral involving the mode propagating in the guiding structure and the mode of the WGDR, both normalized, and a gaussian function of the difference between the longitudinal propagation constant $\beta_{wg}$ in the guiding structure and the transverse propagation constant in the resonator **[15, 16]**. The selective excitation of resonances can then be based on two parameters, namely their spatial extension and their propagation constant. Modes confined in the inner region are favored if the transverse extension of the mode propagating in the guiding structure is not greater than the thickness of the inner region; indeed in this case the overlap integral for modes distributed over the whole structure decreases with the ratio between the thickness of the inner region and the total thickness, respectively. Analogously, the propagation constant $\beta_{wg}$ must be equal to the transverse propagation constant $\beta_{eff}(r)$ of the selected resonance, calculated in the overlap region. The value of $\beta_{eff}(r)$ can be deduced from the relation $e^{in\varphi} = e^{i\beta_{eff}(r)s(r)}$, where $n$ is the azimuthal index of the mode, and $S(r)$ the curvilinear coordinate along a circumference of radius $r$ (see also Ref. 8). The previous analysis can be analogously developed in terms of (local) phase velocities, given by $v_j(r) = \frac{\omega \cdot r}{\beta \cdot r_m} = \frac{\omega}{\beta_{eff}(r)}$ for the resonator (Eq. 5), and by $v_{wg} = \frac{\omega}{\beta_{wg}}$ for the waveguide. From the considerations made in subsection 4a, it follows that an excitation characterized by $v_{wg}$ equal to the phase velocity $v_j(r)$ of modes confined in the inner region of a composite resonator becomes more and more selective when $\varepsilon_1 - \varepsilon_2$ increases.

Different losses in the different regions of a multiple-band WGDR also lead to a discrimination among resonance modes; for instance, if the outer discs are sufficiently lossy, modes distributed over the whole structure exhibit low merit factors and are then difficult to be excited [17].

The above discussion evidences that a selective excitation of high frequency modes of the composite WGDR is allowed by the inhomogeneity of the structure; the inhomogeneity in the real part of the dielectric constant can lead to a differentiation of the coupling coefficients, while the inhomogeneity in its imaginary part can discriminate the merit factors. In this sense, the working principle of a partially coupled WGDR is also based on the inhomogeneity of its dielectric constant.

When only resonance modes confined in the inner region are excited, the fully coupled multiple-band WGDR behaves as a partially coupled one, so that Eqs. 1-3 apply. Analogously, in this condition the analysis of the related merit factors can be developed following the subsection 4a.

## 5. Preliminary experimental results

As a practical rule, when a multiple-band WGDR is designed, first the frequencies of interest must be fixed; the materials forming their different parts can then be chosen with suitable dielectric properties, so the relative maximum and minimum allowed radius follows from Eqs. 1 and 2. The radius $r$ of the structure can be finally chosen among the common allowed values.

As preliminary experimental verification, a fully coupled composite resonator made with high density polyethylene and teflon was investigated in the bands around 20 GHz and 400 GHz, using a Millimeter Vector Network Analyzer (from ABmm, Paris) [18]. Polyethylene exhibits a real dielectric constant $\varepsilon_{PE} = 2.31$, which is essentially flat over all the whole frequency range of interest **[19]**. From measurements of merit factors of polyethylene WGDRs, it follows that **tan $\delta_{PE}$** is better than 6*10$^{-4}$ around 20 GHz and of the order of 3*10$^{-4}$ around 400 GHz. For teflon, **$\varepsilon_{TE}$ = 2.07** can be safely assumed in the considered frequency range **[20]**; its loss angle at 20 GHz has been measured and results about equal to that of polyethylene, while in the high frequency region it is about three times greater. A multiple-band WGDR having outer discs of teflon and a central ring of polyethylene can be then realized. The radius of this resonator, according to the discussion at the beginning of this section, was chosen equal to 30 mm. The thickness **t** of the different parts, chosen in order to sustain few axial modes in the related frequency band, can be evaluated from the condition

$$2 \times \left(1 - \frac{\varepsilon_{out}}{\varepsilon_{int}}\right)^{1/2} \times \frac{t}{\lambda} \gtrsim 1,$$ as discussed in Sect. 4a of Ref **8**; here **$\varepsilon_{int}$** and **$\varepsilon_{out}$** are the dielectric constants delimiting the planar surfaces that confine the resonance mode. It follows that 1.85 mm for the polyethylene ring and 13 mm for the whole structure are convenient thicknesses. To reduce the number of high frequency modes confined in the inner region, the radial thickness of the polyethylene ring is chosen equal to 1.8 mm. In Fig. 3 the resulting section of this composite WGDR is sketched, together with the exciting dielectric waveguides.

The low frequency limit of the proposed resonator can be calculated according to the analysis made in Sect. 4 of **[8]**. In particular, from Eq. 1 it follows that $\nu_{min} = \frac{\omega_{min}}{2\pi} \approx 23\,\text{GHz}$ so that, from Eq. 25 of **[8]**, it can be deduced that $\nu_{min}$ also corresponds to the minimum effective working frequency of the investigated device.

Since the high frequency modes distributed over the whole structure are expected to be hardly excited, owing to the joint effect of a lower overlap integral and a lower merit factor, the upper working frequency $\nu_{upp}$ of the composite WGDR can be calculated considering only the modes confined in the polyethylene ring.

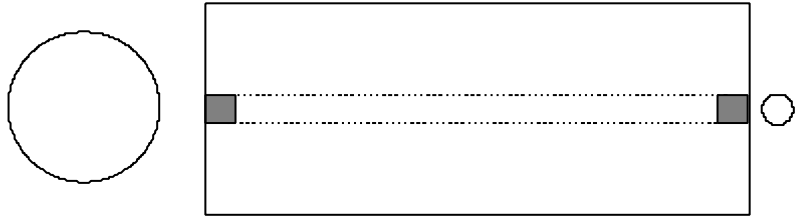

*Fig. 3* *Side view of the employed fully coupled inhomogeneous resonator, together with the low frequency and high frequency waveguides (circles). The shaded area represents the high frequency region.*

By means of the analytical approach developed in **[10]**, it results that around 400 GHz modes with axial indices **l**=0,1 and with radial indices **m** =0,1,2,3 are allowed, both for WGE and WGH family. Starting from Eq. 2, the ideal frequency $n_{max}$, calculated for polyethylene, results of about 3000 GHz. Taking into account the presence of the transverse modes, all of them supposed effectively excited, the maximum allowed frequency is given by $n_{upp} = \frac{n_{max}}{L \cdot M}$, where **L** is the total number of the axial modes and **M** the total number of radial modes, according to the analysis made in Sect. 4a of **[8]**. It follows then $n_{upp} \approx 400\,\text{GHz}$, which is self-consistent with the frequency employed for the calculation of **L** and **M**.

### *5a. Low frequency characterization*

The low frequency characterization of the investigated resonator was made in the frequency range 18-28 GHz. A fused quartz cylindrical rod was used as exciting waveguide in the so-called reaction configuration, the excitation of the WGDR being obtained by means of the evanescent field that surrounds the waveguide **[12]**. A schematic view of the experimental setup is reported in Fig. 4.

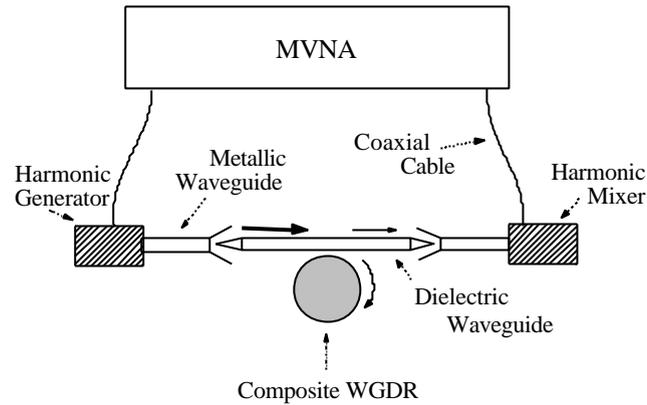

*Fig. 4* *Schematic view of the experimental setup.*

Fig. 5 shows the absorption spectrum of the fundamental $WGE_{n,0,0}$ modes of the resonator. Similar results were obtained for $WGH_{n,0,0}$ modes. The **Q** of the observed resonances rapidly decreases with the frequency, owing to the irradiation losses, being its value around $\mathbf{n_{min}}$ of the order of 50. As expected, also the coupling efficiency decreases with the frequency **[17]**. The critical coupling in Fig. 5 is reached between the resonances labeled 6 and 7, so the amplitude of the next peaks in the absorption curve decreases. For frequencies approaching the lower limit of the considered interval the resonances become unresolved, confirming the estimation of the lower limit frequency of the investigated resonator.

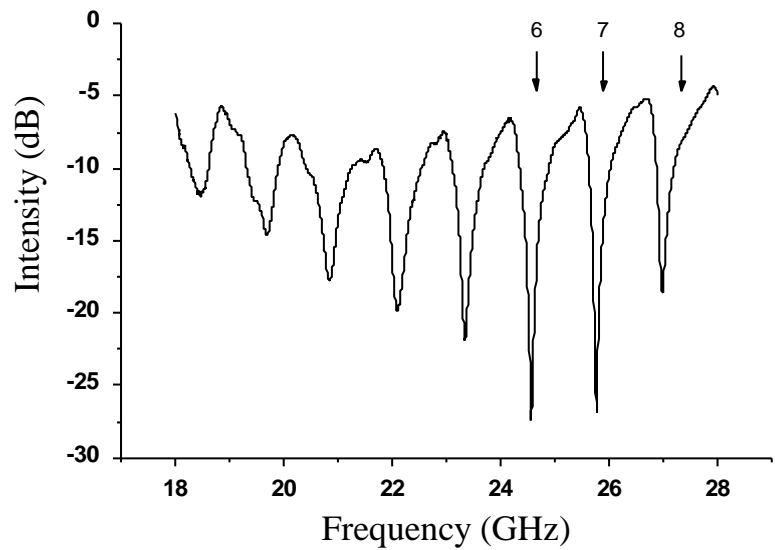

*Fig. 5* *Low frequency absorption spectrum of fundamental $WGE_{n,0,0}$ modes.*

The irradiation losses of planar homogeneous structures can be quantified using the characteristic equation of a cylindrical waveguide in the limit of negligible axial propagation constant, consistently with the approximation made in Sect. 3 of Ref. **[8]**. In this limit the characteristic equation splits in two different equations (one for WGE modes, the other for WGH modes), containing the frequency and the parameters of the resonator; the (complex) frequency solutions give the resonance frequency and the relative merit factor. Fig. 6 reports the calculated and above measured merit factors.

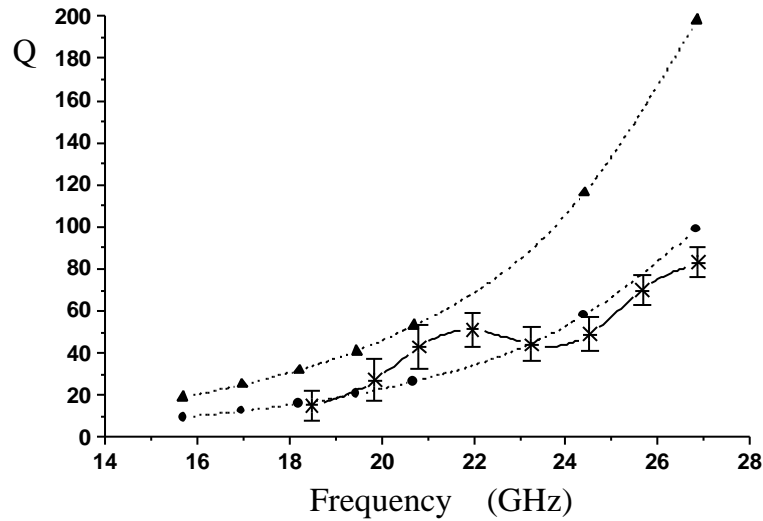

*Fig. 6*    *Q merit factors of the resonances reported in Fig. 5. The triangles indicate the calculated unloaded merit factors, the circles the merit factors calculated at the critical coupling, and the stars the measured values. Interpolation lines are also reported.*

To take into account the effect of the coupling on the merit factor, for simplicity all resonances have been considered critically coupled; more accurate calculations can be done following the modeling described in Ref. **[21]**. The analysis of several other configurations (with different dielectric constant), confirms that the expected **Q** value near $n_{min}$ is of the order of some tens; in addition, the **Q** increase exponentially with the frequency and the rate of increase strongly increases with $e_{int}$.

*5b. High frequency characterization*
The high frequency characterization of the resonator was made using the MVNA, powered with the ESA extension **[18]**.

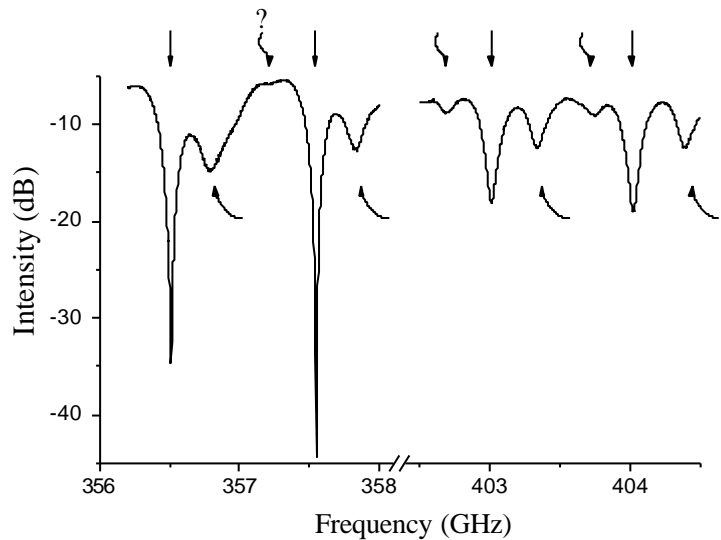

*Fig. 7*  *High-frequency absorption spectra of WGH mode. Identical arrows represent a tentative assignment of modes belonging to the same family.*

The experimental setup, optimized to work till to 450 GHz, is similar to that sketched in Fig. 4, the only significant difference being the reduced diameter of the exciting waveguide, now of the order of 0.5 mm. The investigated frequency interval ranges from about 355 GHz to about 405 GHz.

Fig. 7 reports the absorption spectra of WGH modes obtained around 357 GHz and around 403 GHz, over intervals greater than the free spectral range; similar results were obtained for WGE modes. Several families of resonances are excited; identical arrows in Fig. 7 represent a tentative assignment of modes belonging to the same family. The relative merit factor is of the order of 3000, as expected from the loss angle of polyethylene at these frequencies. The resulting free spectral range is about the same for the different families. Around 403 GHz the spectrum of the composite resonator, although more dense than that obtained around 355 GHz, is still resolved, since only some of the allowed modes are effectively excited.

A full confirmation of the proposed theoretical model was obtained investigating the confinement of the e.m. field in the inner polyethylene ring. To this purpose an additional receiving antenna characterized by a sharp tip **[12]** was moved parallelly to the axis of the resonator (**z** coordinate), near its rim. The field intensity of the resonance at 356.5 GHz is reported in Fig. 8.

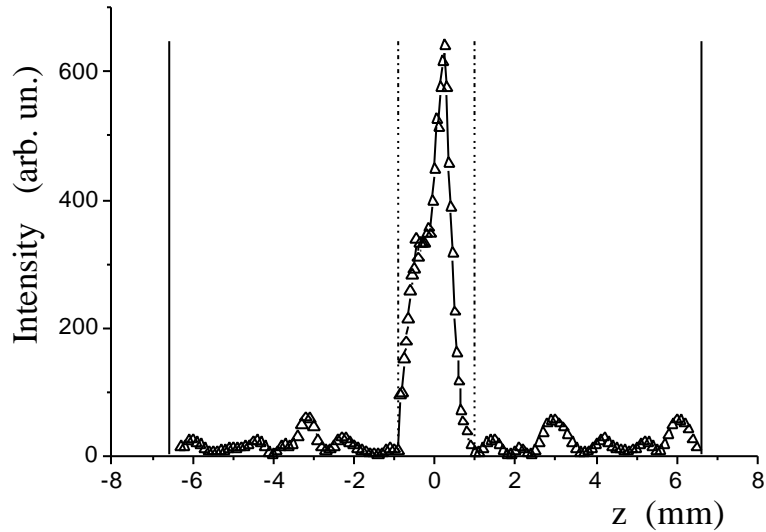

*Fig. 8* *Electromagnetic field intensity along the z-axis of the resonator for the 356.5 GHz resonance, measured near the rim.*

The occurrence of a main peak centered around the inner region (delimited by dashed lines in Fig. 8) confirms that the small difference between the dielectric constant of polyethylene and that one of teflon confines the high frequency mode with **l**=0. The secondary peaks in Fig. 8 represent transverse modes of the whole structure not correlated to the main one, as verified by changing the position of the exciting waveguide. Indeed, while the main peak disappears for excitation outside the central region, the secondary ones are always present. The presence of these peaks also explain the asymmetry observed in the field profile of the main mode.

## 6. Conclusions and possible applications

The realization of an ultra-wideband resonator is an appealing and very promising challenge. In this paper it is proposed and experimentally characterized a very simple composite WGDR whose effective band includes frequencies differing at least by a factor 20. The obtained results substantiate the theory, developed here and in the companion paper, which foresees working frequencies only limited by the dielectric properties of the employed materials; with a suitable choice of these materials, the allowed frequencies can differ by several decades, also for slight inhomogeneity in the dielectric properties of the composite structure.

The peculiar characteristics of the proposed devices make accessible novel experimental methodologies. A sample placed inside the inner region of a composite WGDR can be simultaneously submitted to e.m. fields having very different frequencies; the study of double resonance phenomena or multi-photon transitions is then allowed with all the advantages of an open structure resonator and without its usual band limitations. The characterization of the excited states of a sample represents, for instance, a natural application of multiple-band WGDRs.

A novel attention should be also paid to the media characterized by non-linear electrodynamic properties (see, for instance, Ref. **[22]**) that, used in conjunction with the proposed devices, could lead to the realization of versatile tunable sources. On the other side, the generation of high order harmonics in a non-linear medium experiencing a single low frequency radiation can be optimized employing a composite resonator as a high order filter. The non-linear behavior is also characteristic of a multiple-band WGDR whose inner region includes an active medium for the high frequency radiation, as obtained when the inner disc is a quantum cascade laser **[3]**; the whole composite WGDR can be designed to operate itself as a lower frequency laser excited by an 'intracavity optical pumping'.

Finally, the concept of a dielectric resonator showing a multiple-band spectral response can be generalized increasing the number of nested regions.

**Acknowledgment.** This investigation was partially supported by NATO Grant PST.CLG.976444.

## REFERENCES

1)  J. Yu, X.S. Yao, and L. Maleki, "High-Q whispering gallery mode dielectric resonator bandpass filter with microstrip line coupling and photonic bandgap mode-suppression", IEEE Microwave Guided Wave Lett. vol. 10, pp. 310-312, (2000).
2)  M.E. Tobar, E.N. Ivanov, P. Blondy, D. Cros, and P. Guillon, "High-Q whispering gallery traveling wave resonators for

oscillator frequency stabilization", IEEE Trans. Ultrason. Ferroelectr. Freq. Control, vol.47, no.2, pp.421-426, (2000).

3) C. Gmachl, J. Faist, F. Capasso, C. Sirtori, D.L. Sivco, and A.Y. Cho, "Long-wavelength (9.5-11.5 µm) microdisk quantum-cascade lasers", IEEE J Quantum Elect., vol.33, pp.1567-1573, (1997).

4) V. Sandoghdar, F. Treussart, J. Hare, V. Lefevre-Seguin, J.M. Raimond, and S. Haroche, "Very low threshold whispering-gallery-mode microsphere laser", Phys. Rev. A, vol.54, no.3, pp. R1777-1780, (1996).

5) A.N. Oraevsky, M.O. Scully, T.V. Sarkisyan, and D.K. Bandy, "Using whispering gallery modes in semiconductor microdevices", Laser Phys., vol.9, no.5, pp. 990-1003, (1999).

6) G. Annino, D. Bertolini, M. Cassettari, M. Fittipaldi, I. Longo, and M. Martinelli, "Dielectric properties of materials using whispering gallery dielectric resonators: Experiments and perspectives of ultra-wideband characterization", J. Chem. Phys., vol. 112, no. 5, pp.2308-2314, (2000).

7) G. Annino, M. Cassettari, M. Fittipaldi, I. Longo, M. Martinelli, C.A. Massa, and L.A. Pardi, "High-Field, Multifrequency EPR Spectroscopy Using Whispering Gallery Dielectric resonators", J. Magn. Reson., vol. 143, pp. 88-94, (2000).

8) G. Annino, M. Cassettari, and M. Martinelli, "Study On Planar Whispering Gallery Dielectric Resonators. General Properties", Int. J. Infrared Millim. Waves, this volume.

9) H. Peng, "Study of Whispering Gallery Modes in Double Disk Sapphire Resonators", IEEE Trans. Microwave Theory Tech., vol. 44, pp. 848-853, (1996).

10) G. Annino, M. Cassettari, I. Longo, and M. Martinelli, "Analysis of stacked whispering gallery dielectric resonators for submillimeter ESR spectroscopy", Chem. Phys. Lett., vol. 281, pp. 306-311, (1997).

11) S.C. Hagness, D. Rafizadeh, S.T. Ho, and A. Taflove, "FDTD microcavity simulations: design and experimental realization of waveguide-coupled single-mode ring and whispering-gallery-mode disk resonators", J. Lightwave Technol., vol.15, no.11, pp.2154-2165, (1997).

12) G. Annino, M. Cassettari, I. Longo, and M. Martinelli, "Whispering Gallery Modes in Dielectric Resonator:

Characterization at Millimeter Wavelength", IEEE Trans. Microwave Theory Tech., vol. 45, no. 11, pp. 2025-2034, (1997).

13) M.L. Gorodetsky and V.S. Ilchenko, "Optical microsphere resonators: optimal coupling to high-Q whispering-gallery modes", J. Opt. Soc. Am. B, vol.16, no.1, pp. 147-154, (1999).

14) B.E. Little, J.-P. Laine, and H.A. Haus, "Analytic theory of coupling from tapered fibers and half-blocks into microsphere resonators", J. Lightwave Technol., Vol. 17, pp. 704-715, (1999).

15) H.A. Haus, W.P. Huang, S. Kawakami, and N.A. Whitaker, "Coupled-mode theory of optical waveguides", J. Lightwave Technol., vol LT-5, pp. 16–23, (1987).

16) B.E. Little, S.T. Chu, H.A. Haus, J. Foresi, and J.-P. Laine, "Microring resonator channel dropping filters", J. Lightwave Technol., vol 15, pp. 998–1005, (1997).

17) X. H. Jiao, P. Guillon, and J. Obregon, "Theoretical analysis of the coupling between whispering-gallery dielectric resonator modes and transmission lines", Electron. Lett., Vol 21, pp. 88-89, (1985).

18) P. Goy, M. Gross and J.M. Raimond, in: Proc. 15[th] Int. Conf. On IR and mm Waves, Orlando, Florida, ed. R.J. Temkin (Plenum Press, New York, London, 1990), p. 172, (1990).

19) M.N. Afsar, and K.J. Button, "Millimeter-wave dielectric measurement of materials", Proc. IEEE, Vol. 73, pp. 131–153, (1985).

20) M.N. Afsar, "Precision millimeter-wave measurements of complex refractive index, complex dielectric permittivity, and loss tangent of common polymers", IEEE Trans. Instrum. Meas., vol. IM-36, pp. 530–536, (1987).

21) G. Annino, M. Cassettari, M. Fittipaldi, and M. Martinelli, "Complex response function of whispering gallery dielectric resonators", Int. J. Infrared Millim. Waves vol. 22, pp. 1485-1494 (2001).

22) P.P. Absil, J.V. Hryniewicz, B.E. Little, P.S. Cho, R.A. Wilson, L.G. Joneckis, and P.T. Ho, "Wavelength conversion in GaAs micro-ring resonators", Opt. Lett., vol. 25, pp. 554-556, (2000).